\RequirePackage{ifpdf}
\documentclass{PoS}

\title{Lunar detection of ultra-high-energy cosmic rays and neutrinos with the Square Kilometre Array}
\ShortTitle{Lunar detection of ultra-high-energy cosmic rays and neutrinos with the SKA}

\usepackage[authoryear,sort]{natbib}
\bibpunct{(}{)}{;}{a}{}{,}

\newcommand{\soton}{Univ.\ of Southampton}
\newcommand{\nijmegen}{Radboud Univ.\ Nijmegen}
\newcommand{\astron}{ASTRON}
\newcommand{\karlsruhe}{KIT}
\newcommand{\erlangen}{Univ.\ of Erlangen-Nuremberg}
\newcommand{\groningen}{Univ.\ of Groningen}
\newcommand{\manchester}{Univ.\ of Manchester}
\newcommand{\adelaide}{Univ.\ of Adelaide}
\newcommand{\iowa}{Univ.\ of Iowa}
\newcommand{\atnf}{CSIRO ATNF}
\newcommand{\lebedev}{Lebedev Physical Institute}
\newcommand{\santiago}{Univ.\ de Santiago de Compostela}

\author{
 \speaker{J.D.~Bray}$^1$, J.~Alvarez-Mu\~niz$^2$, S.~Buitink$^3$, R.D.~Dagkesamanskii$^4$, R.D.~Ekers$^5$, H.~Falcke$^{3,6}$, K.G.~Gayley$^7$, T.~Huege$^8$, C.W.~James$^9$, M.~Mevius$^{10}$, R.L.~Mutel$^7$, R.J.~Protheroe$^{11}$, O.~Scholten$^{10}$, R.E.~Spencer$^{12}$ and S.~ter~Veen$^3$\\
 $^1$\soton ;
 $^2$\santiago ;
 $^3$\nijmegen ;
 $^4$\lebedev ;
 $^5$\atnf ;
 $^6$\astron ;
 $^7$\iowa ;
 $^8$\karlsruhe ;
 $^9$\erlangen ;
 $^{10}$\groningen ;
 $^{11}$\adelaide ;
 $^{12}$\manchester \\
 E-mail: \email{j.bray@soton.ac.uk}
}

\abstract{
 The origin of the most energetic particles in nature, the ultra-high-energy (UHE) cosmic rays, is still a mystery.  Only the most energetic of these have sufficiently small angular deflections to be used for directional studies, and their flux is so low that even the 3,000~km$^2$ Pierre Auger detector registers only about 30 cosmic rays per year of these energies.  A method to provide an even larger aperture is to use the lunar Askaryan technique, in which ground-based radio telescopes search for the nanosecond radio flashes produced when a cosmic ray interacts with the Moon's surface.  The technique is also sensitive to UHE neutrinos, which may be produced in the decays of topological defects from the early universe.

 Observations with existing radio telescopes have shown that this technique
is technically feasible, and established the required procedure: the radio
signal should be searched for pulses in real time, compensating for
ionospheric dispersion and filtering out local radio interference, and
candidate events stored for later analysis.  For the Square Kilometre Array
(SKA), this requires the formation of multiple tied-array beams, with high
time resolution, covering the Moon, with either SKA1-LOW or SKA1-MID.  With its large collecting area and broad bandwidth, the SKA will be able to detect the known flux of UHE cosmic rays using the visible lunar surface --- millions of square km --- as the detector, providing sufficient detections of these extremely rare particles to address the mystery of their origin.

}

\FullConference{Advancing Astrophysics with the Square Kilometre Array,\\
		June 8-13, 2014\\
		Giardini Naxos, Sicily, Italy }

\begin{document}

\section{Introduction}

Cosmic rays are extraterrestrial high-energy particles with a spectrum
\citep{CRSpectrum1,CRSpectrum2} ranging from around $10^9$\,eV up to at
least $10^{20}$\,eV.  The most energetic, ultra-high-energy (UHE) cosmic
rays have an incredible amount of energy for a single particle ---
comparable to a well-served tennis ball --- and their origin is unknown.
The usual mechanism invoked for producing cosmic rays is the acceleration
of charged particles by magnetic fields around shocks in the interstellar
medium, via the Fermi process.  However, the ability of a shock or other
region of magnetic turbulence to accelerate a particle up to the highest
energies is limited by its ability to keep the particle confined as it
accelerates \citep{Hillas}, and the observed UHE cosmic rays have energies
around the limit achievable by known astrophysical objects such as active
galactic nuclei, starburst galaxies, and gamma-ray bursts \citep{Sources},
suggesting that other mechanisms may need to be invoked to overcome this
limit \citep{protheroe2003}.  Alternatively, UHE cosmic rays could
originate from the decay of hypothetical supermassive particles, which
could either constitute dark matter, or be produced by topological defects
such as kinks and cusps in cosmic strings formed in the early universe
\citep{Lunardini12,Berezinsky11}.  Identifying the source of UHE cosmic
rays would improve our understanding of the most extreme astrophysical
objects in the universe and the particle acceleration mechanisms they can
support, and potentially establish the existence of hitherto-unknown exotic
particles.


The study of UHE cosmic rays is difficult because of their extremely low
flux: the cosmic ray spectrum follows a power-law of approximately
$E^{-2.7}$, and at the highest energies the flux drops so low that less
than one particle is detected per km$^2$ per century.  Associating them
with specific sources is also difficult because, as charged particles, they
are deflected by intervening magnetic fields, so their direction of arrival
does not correspond to the direction of their source.  Furthermore, at
energies exceeding $\sim 6 \times 10^{19}$~eV, they can interact with
photons of the cosmic microwave background, either through photopion
interactions (for cosmic-ray protons) or photodisintegration (for
cosmic-ray nuclei); this GZK effect \citep{GZK1,GZK2} causes attenuation of
the cosmic-ray flux at the highest energies, preventing many of them from
reaching us.  Studies of the few UHE cosmic rays that reach us have found
them to be weakly correlated with the distribution of matter in the nearby
universe \citep{PAO07C1}, which establishes that they are anisotropic, but
the statistics are not sufficient to clearly link them to a specific class
of source.

The observations proposed herein explore two major avenues for improving
our understanding of the origin of UHE cosmic rays. The first is to observe
more cosmic rays of even higher energy.  As the energy increases, the
magnetic deflection is reduced, and it becomes easier to link a particular
cosmic ray with a specific source; together with an increase in the number
of detected particles, correlation studies will be greatly improved.  Also,
the presence or absence of cosmic rays at higher energies will establish
whether the observed cutoff in the spectrum \citep{PAO08E1,PAO08E2} is due
to the GZK effect, in which case there will be a contribution at higher
energies from sources within the GZK horizon; or from an inherent limit on
the cosmic-ray energy which can be attained by their sources, in which case
the cutoff will be sharp.

The second option is to search for UHE neutrinos, which should be produced
both by UHE cosmic ray sources and in GZK interactions.  Since neutrinos
are uncharged, they travel in straight lines, and point directly back to
their original source.  Exotic-particle-decay models for the origin of UHE
cosmic rays also predict hard neutrino spectra up to extremely high
energies, which would be a clear indication that these models were correct.


Due to the reduced cosmic ray flux at higher energies, and because neutrinos interact only weakly, both these methods benefit from extremely large detectors, significantly greater than even the $3,000$~km$^2$ Pierre Auger Observatory \citep{PAO07C2} and $700$~km$^2$ Telescope Array \citep{abu-zayyad2012b}. The resulting strategy is to remotely monitor a large fraction of the Earth's surface from a high-altitude balloon \citep{gorham2009b} or from space \citep{takahashi2009}.  An alternative approach \citep{dagkesamanskii1989} is to use the Moon as the detector, searching for the radio Askaryan pulse \citep{askaryan1962} produced when a UHE cosmic ray or neutrino interacts in the dense lunar regolith. The visible lunar surface has an area of $19,000,000$~km$^2$; even with limited efficiency and angular acceptance, it still constitutes a very large UHE particle detector.

Since this idea was proposed, a series of experiments has been conducted by
multiple teams with a range of radio telescopes: Parkes
\citep{hankins1996}, Goldstone \citep{gorham2004a}, Kalyazin
\citep{beresnyak2005}, the ATCA \citep{james2010}, Lovell
\citep{spencer2010}, the EVLA \citep{jaeger2010}, Westerbork
\citep{buitink2010}, Parkes again \citep{bray2013b}, and Parkes and the
ATCA in combination \citep{bray2011c}, with preparatory work under way with
LOFAR \citep{singh2012}.  These experiments have overcome a range of
technical challenges specific to this type of experiment and, together with
associated theoretical work, have established the viability of this
technique for detecting UHE particles.  The upcoming Square Kilometre Array
(SKA) is an international radio telescope that will have greater
sensitivity than any previous instruments.  With SKA1 it will be able to
achieve the first lunar-target particle detection, while SKA2
will be a powerful instrument for addressing the mystery of
the origin of cosmic rays.

\section{Particle astronomy with a radio telescope}
\label{sec:observations}

In this section, we describe the physical mechanism --- the Askaryan effect --- by which the interaction of high-energy particles produces radio emission, and the methods by which a radio telescope such as the SKA could detect that emission.

\subsection{Particle cascades and radio emission}

\begin{figure}
 \begin{center}
  \begin{minipage}[c]{0.44\linewidth}
   \centering
   \includegraphics[width=\textwidth]{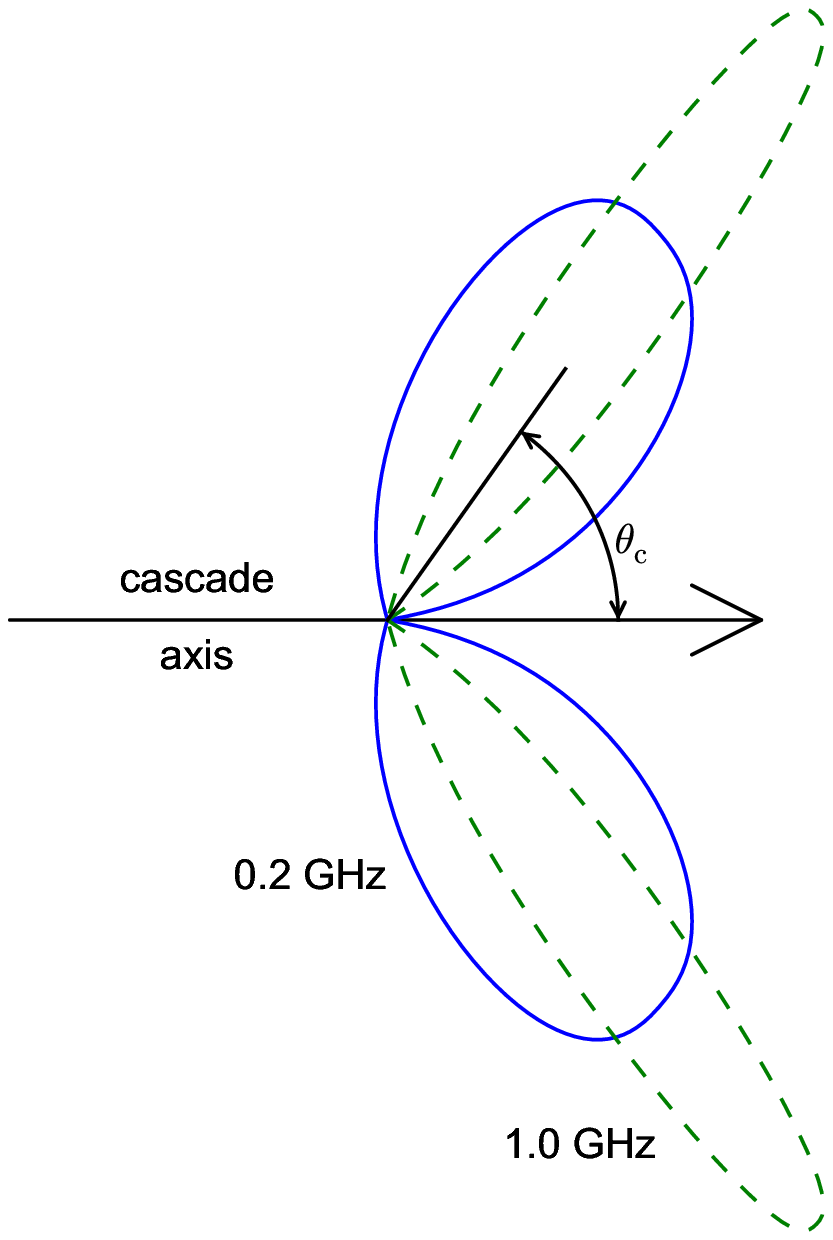}
  \end{minipage}
  \begin{minipage}[c]{0.54\linewidth}
   \centering
   \includegraphics[width=\textwidth]{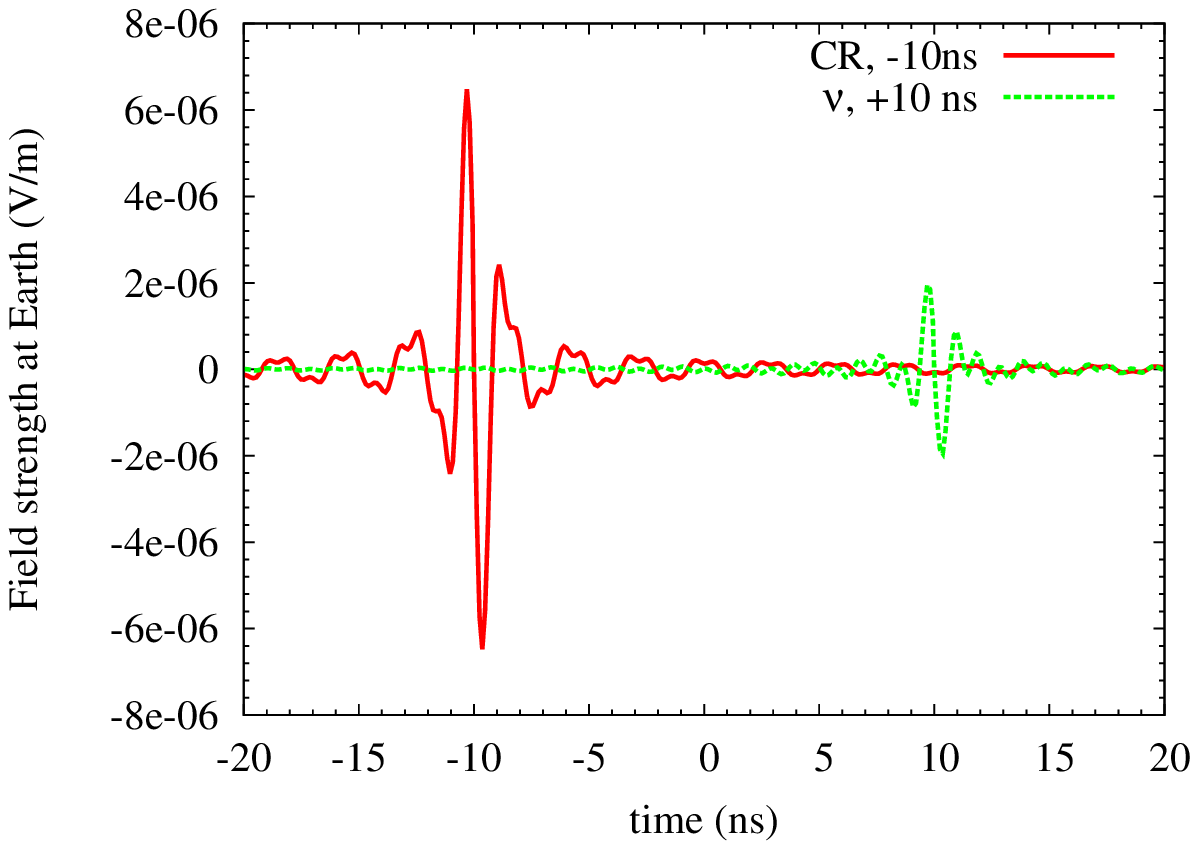}
  \end{minipage}
 \end{center}
 \caption{Left: Askaryan emission from a hadronic particle cascade in regolith.  The emission is directed forward along the cascade axis as a hollow cone at the Cherenkov angle $\theta_{\rm c}$.  The radiation pattern depends on the frequency, as shown. Right: time-domain pulses (excluding dispersive effects) from $10^{20}$~eV cosmic ray (CR) and neutrino ($\nu$) interactions with typical geometries, as might be recorded at Earth over a $350$~MHz to $1.05$~GHz bandwidth.  The pulses have been offset by $\pm 10$~ns for display purposes.}
 \label{fig:askaryan_emission}
\end{figure}

A UHE particle interacting in a dense medium will initiate a particle cascade, with the total energy being distributed over an increasing number of particles as the cascade progresses.  Due primarily to entrainment of electrons in the medium, the cascade develops a net negative charge \citep{askaryan1962}, causing it to radiate coherently at wavelengths greater than the width of the cascade ($\lambda \gtrsim 10$~cm; frequencies up to a few GHz).  This radiation is beamed forward as a hollow cone around the cascade axis at the Cherenkov angle ($\sim{55}^\circ$), at which it manifests as a single pulse with a duration given by the inverse bandwidth (\mbox{$\lesssim 1$}~ns).  For directions away from the Cherenkov angle, the pulse is weaker, and more extended in time.  The emission is more broadly beamed at lower frequencies, with the radiation pattern extending to perpendicular to the cascade axis \citep{NuMoonSim}, and stronger at higher frequencies, as shown in Fig.~\ref{fig:askaryan_emission} (left).

If the primary particle is a cosmic ray, then the cascade consists predominantly of hadrons, and develops (in a dense medium such as the lunar regolith) quite quickly, reaching a length of only \mbox{$\sim 3$}~m \citep{Alvarez-Muniz06}.  If the primary particle is a neutrino, it initiates a hadronic cascade carrying \mbox{$\sim 20$}\% of the primary energy, which behaves in a similar fashion.  Depending on the neutrino flavour and interaction type, the remaining energy may go into an electromagnetic cascade, containing only electrons/positrons and photons, but at ultra-high energies these cascades are highly elongated, suppressing their radio emission, and are not generally detectable \citep{JamesProtheroe09a}.

A cosmic ray interacts as soon as it enters the regolith, so all such cascades occur close to the lunar surface.  At higher frequencies, where the radiation is strongly beamed forward, and refracted further forward as it escapes the lunar surface, the cascade is only detectable if the cosmic ray interacts in a shallow, skimming trajectory, allowing the radiation to escape rather than being directed downward into the Moon.  At lower frequencies, where emission sideways from the cascade is coherent \citep{Veen10}, detection is possible over a greater range of interaction geometries.

Neutrinos interact less strongly than do cosmic rays, and may occasionally pass through the bulk of the Moon and interact while steeply upgoing relative to the local surface, but at ultra-high energies their interaction cross-section is increased, and they are most likely instead to be detected in a skimming trajectory similar to cosmic rays.  They may, however, be well below the lunar surface when they interact, and may be detected at depth up to a few times the radio attenuation length in the lunar regolith (a few tens of metres, depending on frequency).  Because many of them will interact too deeply in the Moon to be detected, neutrinos are less likely to be detected than are cosmic rays \citep{jeong2012}.

The radiated pulse manifests as a transient bipolar fluctuation in the electric field measured by a radio receiver, as seen in Fig.~\ref{fig:askaryan_emission} (right), and confirmed in laboratory measurements at SLAC \citep[see e.g.][]{miocinovic2006}.  As the components of the pulse at different frequencies combine coherently in the voltage domain, the signal-to-noise ratio in power scales linearly with bandwidth, in contrast to the square-root law typical in other fields of radio astronomy.  The sensitivity of a radio telescope defines a minimum detectable pulse amplitude, which in turn defines a minimum detectable particle energy; the SKA, with its superior sensitivity, will probe down to lower particle energies than has previously been possible with this technique. As the pulse has a spectrum which increases in amplitude with increasing frequency, up to an angle-dependent turnover at a few GHz or below due to either decoherence or absorption, observations aiming to minimise the detection threshold tend to favour higher frequencies. Above the threshold energy, the sensitivity of an experiment to a particle flux is defined by the geometric aperture, with units of area times solid angle.  The radio pulse at low frequencies is more broadly beamed, covering a larger solid angle, which favours lower frequencies for observations significantly above the detection threshold. 

This complex dependence of sensitivity on frequency means that both SKA1-LOW and SKA1-MID are useful and complementary in this role, and explore different regions of the energy-flux phase space.  (SKA1-SUR, despite its larger field of view, offers no advantage over SKA1-MID, as the field of view of the latter is sufficient to see the entire Moon.)  The discussion below, except where noted, applies to both instruments.

\subsection{Observational requirements}
\label{sec:requirements}

\begin{figure}
 \begin{center}
  \includegraphics[width=0.55\textwidth]{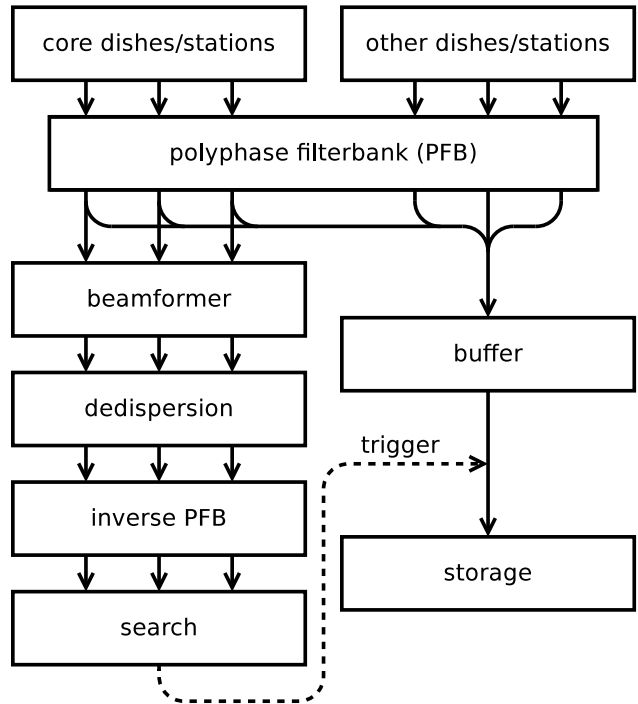}
  \caption{Proposed signal path as described in
Sec.~\protect\ref{fig:sigpath}.  The signals from the core dishes
(SKA1-MID) or stations (SKA1-LOW) are, after being channelised by the polyphase filterbank, formed into sufficient beams to tile across the visible surface of the Moon.  While still channelised, they are dedispersed to compensate for the effects of the ionosphere, and then converted back to pure time-domain signals by an inverse polyphase filterbank.  The resulting signals are then searched searched for a nanosecond-scale pulse meeting several RFI-rejection criteria.  If one is found, it is used to trigger the storage of buffered data from all dishes/stations, which can be retrospectively processed to analyse the pulse with the full sensitivity of the entire array.}
  \label{fig:sigpath}
 \end{center}
\end{figure}

The observational requirements for detecting a nanosecond-scale pulse from the Moon are different from those of other areas of radio astronomy.  Such a short pulse is only detectable in the original time-domain voltages sampled by the receiver, at their full native time resolution.  Storing this quantity of data is usually impractical, so it is generally necessary to maintain a buffer of data and search for a pulse in real time, triggering the storage of the buffer only when a candidate event is found; the proposed signal path is shown in Fig.~\ref{fig:sigpath}.  The trigger criteria need not be perfectly optimised in real time, but they must be good enough that all plausible events are stored so that they can be analysed retrospectively.

As an array, the SKA is only fully sensitive to a transient signal when it
is combined to form a tied-array beam.  Buffered data can (and should) be
kept from individual dishes/stations of the SKA so that they can be
combined retrospectively, but for triggering purposes it is necessary to
form beams with at least the core of the array in real time, to have
sufficient sensitivity to detect candidate events.  The specifications for
SKA1-MID include such a beamformer \citep{SKA_baseline_design}, and a
proposal for a similar feature for SKA1-LOW is under consideration
\citep{stappers_ecp}.  The SKA1-MID beamformer, for bands 1 and~2, will
have the capacity to form sufficient beams to tile across the visible
surface of the Moon (\mbox{$\sim 0.2$} deg$^2$), and to tile in frequency
to cover the entire band.

For a nanosecond-scale pulse to be detected, the signal for each beam should consist of a single sequence of voltage samples, with inverse-bandwidth time resolution, so that a pulse --- which might comprise a single high-amplitude sample --- can be identified.
This presents a problem, as the beamformer acts on the signal after it has passed through a polyphase filterbank, which separates it into frequency channels with inverse-channelwidth time resolution.  Fortunately, it is possible to invert a polyphase filterbank to restore the time-domain signal for each beam, with a minor loss of efficiency \citep{singh2012}.

A lunar-origin nanosecond-scale radio pulse is dispersed as it passes through the ionosphere.  The degree of dispersion is minor compared to that for interstellar signals but, due to the extremely short pulse, it still causes a significant loss of amplitude, and must be corrected.  Because the signal is channelised during the beamforming stage, it is relatively simple to dedisperse it by applying an appropriate phase factor, but it may be necessary to test the signal for a trigger condition at multiple dispersion measures, depending on how precisely the dispersion can be predicted.  Current techniques are based on ionosonde measurements or GPS-based global electron content maps; improvements are possible through direct measurements of dispersion to GPS satellites from a ground station at the telescope site (currently on trial at Parkes) or through measurements of the Faraday rotation of polarised thermal emission from the Moon itself \citep{mcfadden2011}.

As transient detection experiments aim to detect single, unrepeated events, they are highly susceptible to radio-frequency interference (RFI): it is necessary to reliably exclude every single instance of nanosecond-scale impulsive RFI to be certain that any remaining events are true lunar-origin Askaryan pulses.  The simplest way to achieve this is to require a coincident detection on multiple independent antennas \citep[e.g.][]{gorham2004a}, but this limits the sensitivity of the experiment.  In experiments which have used the full coherent sensitivity of a telescope's entire collecting area, the most effective technique is to exclude pulses that are detected with multiple beams directed at different parts of the Moon, as employed with the WSRT \citep{buitink2010} and the Parkes telescope \citep{bray2013b}.  In the latter case, RFI was successfully excluded with high fidelity in real time, and completely in retrospective processing.  The SKA will have the further advantages of a less active RFI environment and a greater number of beams on the Moon, and we are confident that RFI triggers can be excluded with sufficient reliability that they will not dominate the real-time trigger rate, and those which do cause a trigger can be eliminated in retrospective processing.

If RFI is excluded, then the trigger rate will be dominated by thermal fluctuations in the voltage.  As each trigger causes the current contents of the buffers to be stored to disk, the speed with which these data can be stored will determine the supportable trigger rate, and hence how low the trigger threshold can be set.  Assuming a trigger condition of a global OR over all the on-Moon beams, so that a pulse can be detected from anywhere on the Moon, a trigger threshold of \mbox{$\sim 7\sigma$} (relative to the thermal noise) would suffice to keep the background trigger rate around 1~Hz.  Assuming a buffer length of 10~$\mu$s, primarily to assist in characterising RFI, and 8-bit sampling, this results in a manageable data-storage rate of \mbox{$\sim 10$}~MB/s.  This value applies to both SKA1-LOW and SKA1-MID: the former has more stations and the latter has a higher sampling rate, but these effects approximately cancel out.

Retrospective analysis will achieve greater sensitivity than the real-time beams, due to the use of the entire array rather than just the beamformed core, resynthesising the beam directly on the source position, more precise dedispersion, etc.  These effects are typically of the order of a few tens of percent.  Combined, they might allow retrospective analysis to reach a sensitivity \mbox{$\sim 2\times$} that available in real time; i.e.\ the real-time sensitivity would be worse than the theoretical sensitivity by a factor of 2, or $\sqrt{2}$ in the voltage domain.  This means that a 7$\sigma$ threshold in the real-time trigger would ensure that any pulse exceeding $7\sqrt{2}\sigma \sim 10\sigma$ with the full theoretical sensitivity would be stored and detected.  We use this 10$\sigma$ threshold when calculating the sensitivity in Sec.\ \ref{sec:performance}.  At this level, the false detection rate is less than one per century.

\section{SKA performance}
\label{sec:performance}

Simulations of the SKA sensitivity to UHE particles using the lunar
Askaryan technique were performed by a detailed Monte Carlo code
\citep{JamesProtheroe09a}. This code simulates particle propagation and
interactions in the Moon, radio-wave production according to
\citet{Alvarez-Muniz06}, its subsequent absorption in and transmission
through lunar rock, and signal detection based upon an assumed instrumental
sensitivity and the calculated pulse strength. The accuracy of this
simulation has been verified by comparisons with (semi-)analytic
calculations in limiting cases \citep{gayley2009,NuMoonSim}.

The dominant uncertainties in the simulation are the interaction
cross-sections for UHE neutrinos and the effects of lunar surface
roughness.  The former have an uncertainty at $10^{20}$~eV of \mbox{$\sim
20\%$} within the standard model \citep{cross-sections}, and exotic
extra-dimensional models suggest that they could be larger by 1--2 orders
of magnitude \citep{connolly2011}; we use the standard-model cross-sections
of \citet{gandhi}.  For the lunar surface, we use the self-affine model of
\citet{shepard1995}, implementing it by randomly deviating the lunar
surface on length scales corresponding to the characteristic size of the
radiation transmission region on the surface.

Due to the broadband nature of the signal, the optimal frequency range for
lunar observations with the SKA can only be determined by simulating all
frequency bands up to the highest-frequency band for which beamforming
across the entire Moon is practical.  We therefore consider the SKA1-LOW
band as well as the two lowest-frequency receivers for SKA1-MID (SKA1-MID 1
and~2). As described in Sec.~\ref{sec:observations}, the sensitivity of
these SKA bands to lunar signals must be calculated relative to the lunar
thermal emission which, at approximately $225$~K
\citep{troitskijtikhonova70}, is brighter than the sky at frequencies above
approximately $168$~MHz. Averaged over the SKA1-LOW band, both the lunar
emission and the sky background will have similar strengths, so the SKA1-LOW
sensitivity from \citet{SKA_baseline_design} is used unchanged. In the case
of SKA1-MID, the lunar emission will dominate, and is added to the system
temperature. The resulting sensitivities are given in
Table~\ref{table:sensitivities}, where the signal detection thresholds
correspond to $10$ times the effective beam root-mean-square noise voltage
$\sigma$, under the assumption that the observational requirements ---
dedispersion, triggering, RFI discrimination, etc.\ --- are satisfied.

\begin{table*}
\begin{center}
\begin{tabular}{c|cc|cc|cc}
\multicolumn{3}{c|}{} & \multicolumn{2}{c|}{SKA1} & \multicolumn{2}{c}{SKA2} \\
\hline
& $f_{\rm min} $& $f_{\rm max} $ & $A_{\rm eff}/T_{\rm beam}$  & $ E_{\rm thresh} $ & $A_{\rm eff}/T_{\rm beam}$  & $ E_{\rm thresh} $\\
Band & (MHz) & (MHz) & m$^2$/K  & V/m/MHz & m$^2$/K  & V/m/MHz \\
\hline
LOW     & 100 & 350 & 1000 & $1.4 \times 10^{-9}$ & 4000 & $7.2 \times 10^{-10}$ \\
MID 1   & 350 & 1050 & 143 & $2.3 \times 10^{-9}$ & 844  & $9.4 \times 10^{-10}$ \\
MID 2   & 950 & 1760 & 143 & $2.1 \times 10^{-9}$ & 844  & $8.7 \times 10^{-10}$ \\
\end{tabular}
\caption{Estimated sensitivities ($A_{\rm eff}/T_{\rm beam}$, including
lunar thermal emission) and signal detection threshold thresholds (V/m/MHz)
for SKA observation bands. SKA1-MID treats both 64 MeerKAT and 190 SKA antennas identically, while SKA2 uses 1,500 SKA antennas.} \label{table:sensitivities}
\end{center}
\end{table*}

We perform these simulations with a simple 10$\sigma$ threshold, and with
additional restrictions on the pulse amplitude, polarisation and point of
origin on the Moon, as described below.  Based on the results, we determine
the effective aperture to cosmic rays and neutrinos, and the resolution in
the energy and arrival direction of a detected cosmic ray.  We neglect any
constraints on the interaction geometry that could be obtained by measuring
the radio pulse spectrum, or (for SKA2) the radiation pattern over the
Earth, which would further improve the resolutions shown here.  The
prospects for determining the mass of a detected cosmic ray \citep[as in
e.g.][]{abraham2010b} are effectively nil.

\subsection{Aperture}

\begin{figure*}
 \begin{center}
  \includegraphics[width=\textwidth]{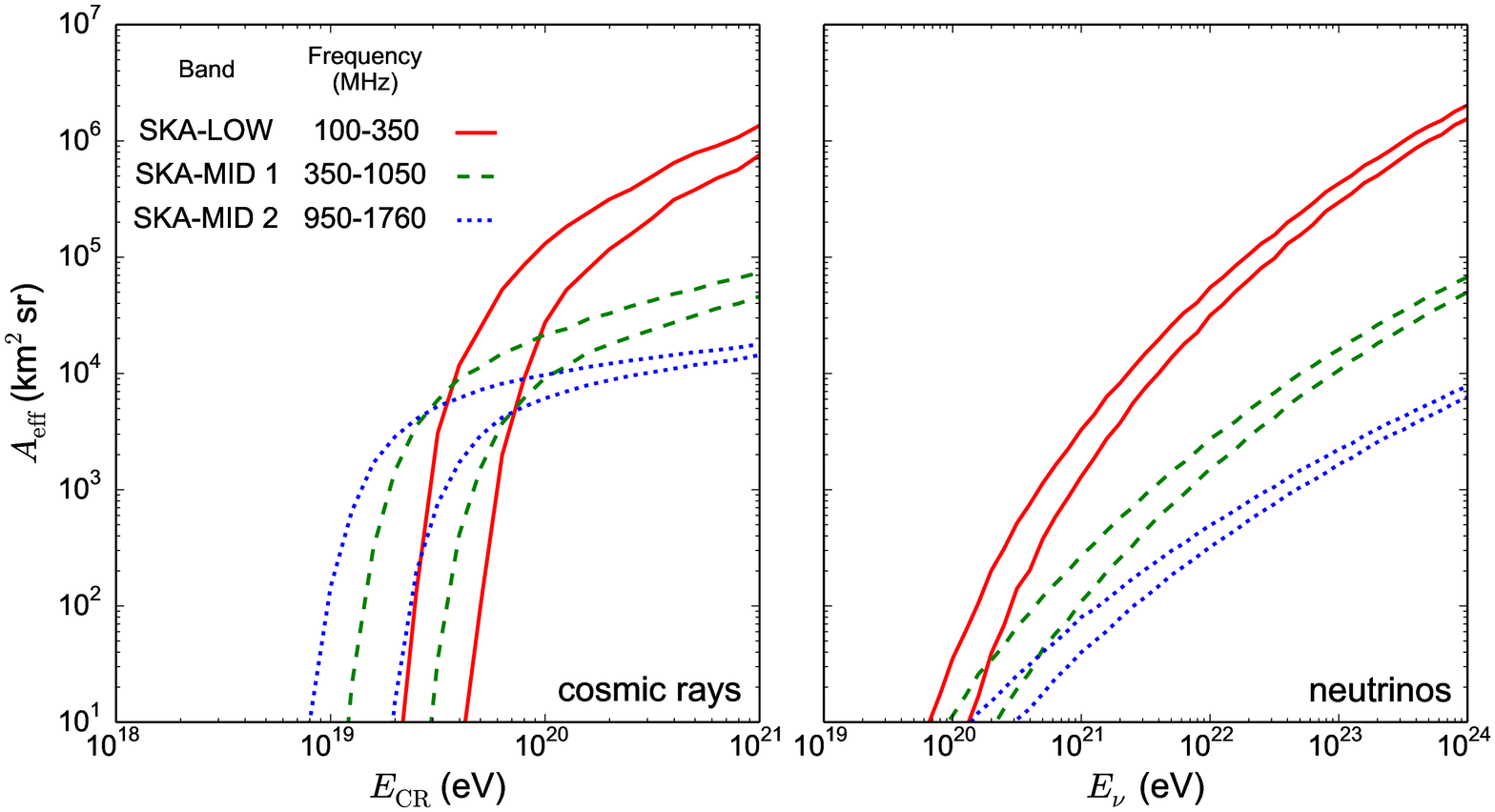}
  \caption{Effective apertures $A_{\rm eff}$ of the SKA to cosmic rays
(left) and neutrinos (right).  In each case, SKA1 is given by the lower
curves, and SKA2 by the upper curves.  Estimates for SKA1-LOW and SKA1-MID are shown, indicating the frequency range of each band.}
  \label{fig:apertures}
 \end{center}
\end{figure*}

The geometric apertures for detection of cosmic rays and neutrinos are
shown in Fig.~\ref{fig:apertures}.  The general trends are as expected from
previous work: observing at lower frequency results in a larger aperture,
due to the larger beaming angle, while observing at higher frequency
results in sensitivity down to lower energies, due to the more intense
radio emission.  Unlike cosmic rays, neutrinos may interact deep in the
lunar regolith, which preferentially attenuates higher-frequency radio
emission, so lower frequencies are more strongly prefered for the detection
of neutrinos.

Fig.~\ref{fig:energres} (left) shows the expected energy-dependent
detection rates for cosmic rays, based on these apertures and an empirical
parameterisation of the UHE cosmic ray spectrum from
\citet{AbrahamAuger2010}.  The expected detection rates for SKA2 are much
greater than those for SKA1: SKA2 at low frequencies would detect 165 cosmic rays above
the 56~EeV threshold used for anisotropy studies by \citet{PAO07C2} per
year of observing time, although such a long exposure would require a
parallel beamforming capacity to allow commensal observations.

\subsection{Energy resolution}
\label{sec:energres}

The interaction of a UHE particle can be observed as a radio pulse with an
amplitude up to some energy-dependent maximum value, which is attained
under an optimal observing geometry.  A detected pulse therefore
establishes a minimum bound on the energy of the original particle.  A soft
upper bound is placed by the steep cosmic ray spectrum: a low-amplitude
pulse is more likely to originate from a lower-energy particle than from a
rarer higher-energy particle observed in a suboptimal geometry.  These two
bounds allow us to establish an effective energy resolution for detected
cosmic rays \citep[similar to][]{scholten2008}.

\begin{figure}
 \begin{center}
  \includegraphics[width=\linewidth]{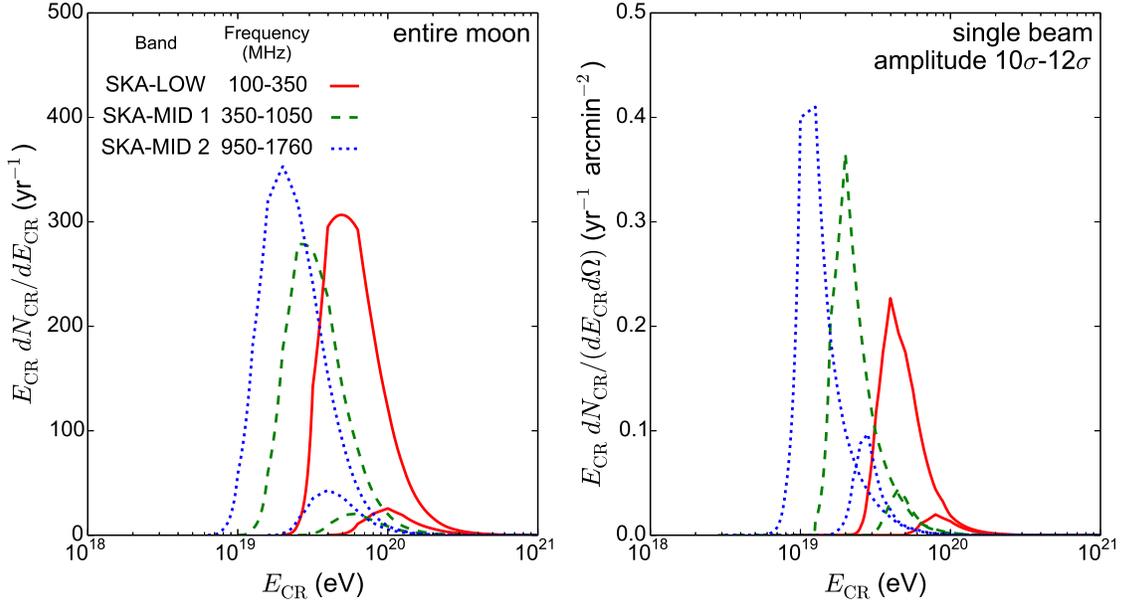}
  \caption{Expected energy distribution of detected cosmic rays, assuming
the parameterised cosmic ray spectrum from \citet{AbrahamAuger2010}, for
(left) coverage of the entire Moon, using the apertures from
Fig.~\protect\ref{fig:apertures}, and (right) for a single beam placed 0.97
lunar radii from the centre of the Moon, requiring detected events to have
an amplitude in the range 10$\sigma$--12$\sigma$ and polarisation within
5$^\circ$ of radial to the Moon.  The latter curve is also scaled by the
beam size, and acts as a measure of the energy resolution attainable with
these constraints on the position, amplitude and polarisation of the pulse.
As in Fig.~\protect\ref{fig:apertures}, the lower curves correspond to
SKA1, and the upper curves correspond to SKA2.  With the scaling and normalisation used here, the number of detected cosmic rays is proportional to the area under the curve.}
  \label{fig:energres}
 \end{center}
\end{figure}

Fig.~\ref{fig:energres} (right) shows the expected energy distribution of
detected cosmic rays for all pulses in the range 10$\sigma$--12$\sigma$,
with further constrains on location and polarisation (see
Sec.~\ref{sec:angres}).  This effectively shows the expected posterior
distribution in energy $E$ for a cosmic ray detected with an amplitude
uncertainty of \mbox{$\pm 1\sigma$}.  Taking the root-mean-square variation
in $\ln E$ gives values in the range 0.27--0.37 for the different frequency
bands, with little variation from this with the different flux models
considered in Sec.~\ref{sec:verification}.  This corresponds to an
uncertainty factor in $E$ of 1.31--1.45, which provides an indicative
figure for the energy resolution.  Note that, since the derivation of this
value assumes the cosmic ray spectrum to be known, it is not a true energy
resolution suitable for detecting spectral features, but it may still be
used for selecting events for directional studies (see
Sec.~\ref{sec:sources}).

\subsection{Angular resolution}
\label{sec:angres}

\begin{figure}
 \begin{center}
  \includegraphics[width=0.48\linewidth]{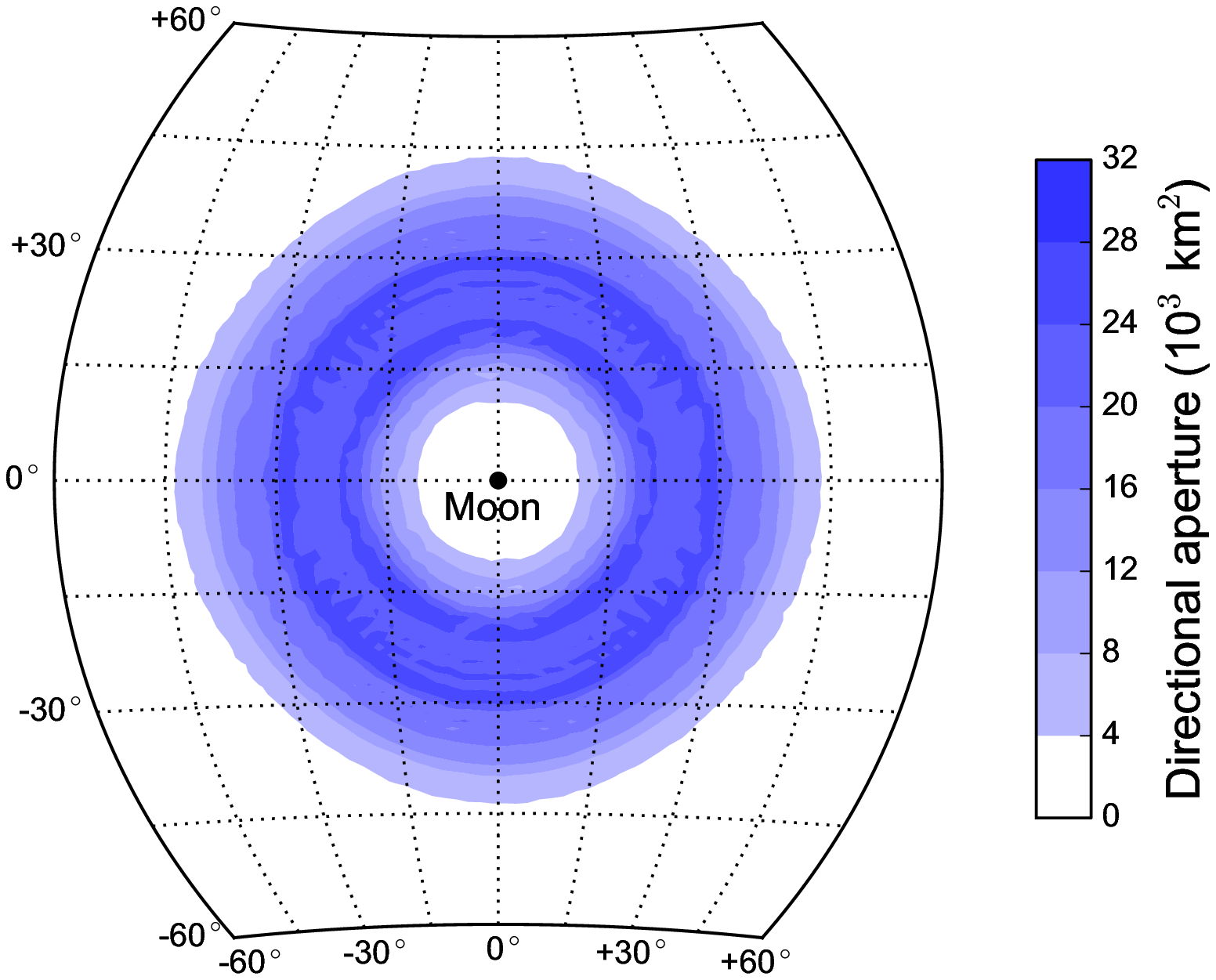}
  \includegraphics[width=0.48\linewidth]{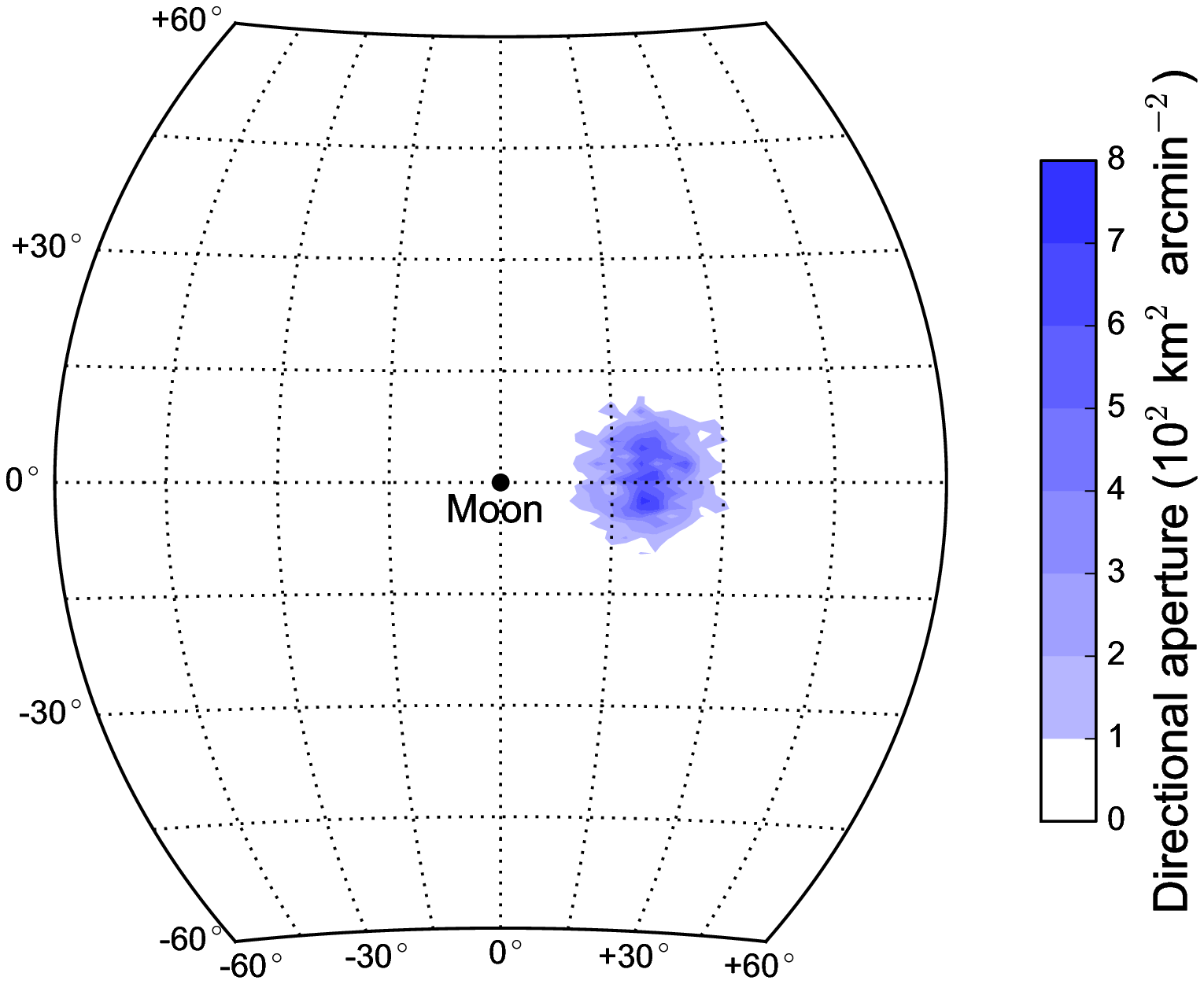}
  \caption{Directional aperture at $10^{20}$~eV to cosmic rays around the Moon with SKA1-LOW for (left) coverage of the entire Moon, and (right) for a single beam placed 0.97 lunar radii from the centre of the Moon, with the polarisation required to be within 5$^\circ$ of radial to the Moon, scaled by the beam size.  The former indicates the field of view, while the latter indicates the angular resolution in the cosmic ray arrival direction, given these constraints on the detected pulse.}
  \label{fig:maps}
 \end{center}
\end{figure}

The field of view from which a cosmic ray can be detected is an annular
zone around the Moon, as shown in Fig.~\ref{fig:maps} (left).  The origin
of a cosmic ray within this zone can be determined from the point of origin
of the pulse on the Moon and the polarisation of the pulse, which is linear
and radial to the particle cascade.  We perform simulations with
restrictions placed on these characteristics in order to determine the
angular resolution with which this origin can be determined.

We assume the origin of a detected pulse to be localised with an effective
baseline of 10~km, a diameter which contains 90--95\% of the antennas of
SKA1-LOW or SKA1-MID.  In practice, for a 10$\sigma$ pulse, it should be
possible to do better than this by measuring the phase slope across the
array, but further improvements have only a small marginal benefit.  For
the purposes of the simulation, based on this baseline and the central
frequency of each band, we require the pulse to originate within a disk of
radius 0.55~arcmin for SKA1-LOW, 0.17~arcmin for SKA1-MID band~1, or
0.08~arcmin for SKA1-MID band~2.  Detectable pulses are expected to be
clustered near the limb of the Moon; for our simulation, we place this disk
0.97 lunar radii from the centre of the Moon.

The polarisation of a pulse can be measured with a precision of
1/$n_\sigma$~radians, where \mbox{$n_\sigma \geq 10$} is the pulse
amplitude.  In our simulation we require the pulse polarisation to fall
within the range \mbox{$\pm 5^\circ$}, corresponding to a pulse near the
detection threshold and neglecting the improved precision possible with
pulses of larger amplitude.  We centre the accepted polarisation range on
an orientation radial to the Moon, which is the expected polarisation for a
majority of detected pulses.

\begin{figure}
 \begin{center}
  \includegraphics[width=0.6\linewidth]{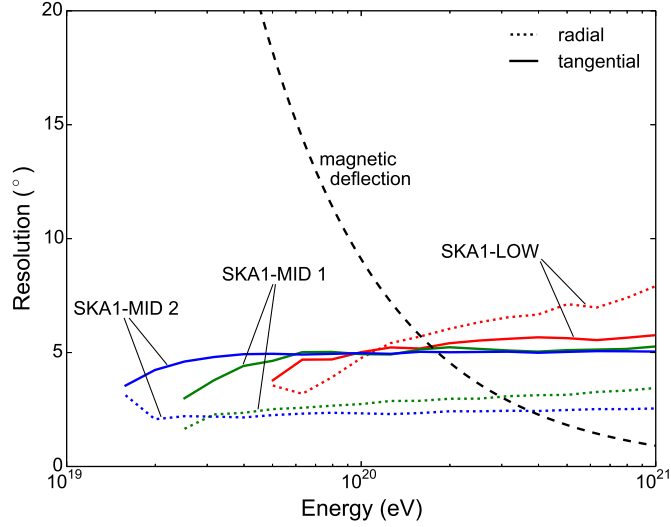}
  \caption{Angular resolution in cosmic ray arrival direction, calculated
as the root-mean-square variation of the aperture distribution as in
Fig.~\protect\ref{fig:maps} (right), in the directions radial and
tangential to the Moon.  The results shown are for SKA1, but the results
for SKA2 for the corresponding bands are similar.  The dashed line corresponds to the derived deflection of cosmic rays in magnetic fields if the observed large-scale anisotropies in cosmic ray arrival directions are due to individual nearby sources; see Sec.~\protect\ref{sec:sources} for details.  At the energy of most expected detections (see Fig.~\protect\ref{fig:energres}, left), this magnetic deflection dominates the angular resolution.}
  \label{fig:angres}
 \end{center}
\end{figure}

With the above constraints on the point of origin and polarisation angle of
the pulse, our simulation produces directional apertures as shown in
Fig.~\ref{fig:maps} (right).  In Fig.~\ref{fig:angres}, we show the
root-mean-square variation of the directional aperture around the mean, in
directions radial and tangential to the Moon.  For comparison, we also show
an estimate of the magnetic deflection of cosmic rays, based on the
localised excesses seen by previous experiments (see
Sec.~\ref{sec:sources}).

\section{Initial prospects with SKA1}
\label{sec:prospects}

Here we consider the objectives that could be addressed with 1,000 hours of
lunar observations with SKA1.  The primary objective of these initial observations would be to verify the lunar Askaryan technique by successfully detecting a lunar particle cascade, but there are also prospects to study the spectrum and anisotropy of UHE cosmic rays, and to search for UHE neutrinos.

\subsection{Verification of the technique}
\label{sec:verification}

The prospects for a successful detection depend on the UHE cosmic ray spectrum, which is not well constrained.  Table~\ref{tab:cr_rates} shows the expected number of detections in 1,000 hours of observations for different theoretically-motivated models of the UHE cosmic ray spectrum from \citet{taylor2011}.  SKA1-MID~2 would most reliably detect at least one cosmic ray, with 4.4~events expected in the most pessimistic case, or a 99\% chance of one or more detections.

\begin{table*}
 \begin{center}
  \begin{tabular}{c|ccc|ccc}
   & \multicolumn{3}{c|}{Scenario A (100\% Fe)} & \multicolumn{3}{c}{Scenario B (100\% Si)} \\
   & \multicolumn{3}{c|}{$E_{\rm max} = 10^{21}$~eV, $\alpha = 1.8$} & \multicolumn{3}{c}{$E_{\rm max} = 10^{20.5}$~eV, $\alpha = 1.6$} \\
   \hline
   $L_{\rm min} =$ && 0~Mpc & 27~Mpc && 0~Mpc & 27~Mpc \\
   \hline
   LOW   && 9.4 & 5.1 && 2.8 & 0.7 \\
   MID 1 && 4.9 & 3.8 && 2.6 & 1.6 \\
   MID 2 && 7.8 & 6.8 && 5.4 & 4.4 \\
  \end{tabular}
  \caption{Expected numbers of cosmic rays detected in 1,000 hours of observations with SKA1, under different scenarios considered by \citet{taylor2011}: for sources emitting a pure iron or silicon composition, with best-fit values for the cutoff energy $E_{\rm max}$ and spectral index $\alpha$.  A local void of radius $L_{\rm min}$ is assumed to exist in the distribution of cosmic ray sources; \mbox{$L_{\rm min} = 0$} corresponds to a continuous distribution.  SKA1-LOW shows the most potential for discriminating between scenarios, while SKA1-MID band~2 is most reliable for detecting at least one event, to allow verification of the technique.}
  \label{tab:cr_rates}
 \end{center}
\end{table*}

\subsection{Measuring the cosmic ray spectrum}
\label{sec:spectrum}

As seen in Table~\ref{tab:cr_rates}, the number of expected cosmic ray detections depends on the assumed model for the cosmic ray spectrum, which offers the prospect of discriminating between these models.  The greatest discriminating power is provided by SKA1-LOW: neglecting the case of a local void in cosmic ray sources, SKA1-LOW would detect an expected 9.4~events in scenario~A and 2.8~events in scenario~B.  These are not an exhaustive sample of possible models, and synergies with other cosmic-ray experiments are complicated by the requirement to establish a common energy scale, but this indicates that event counts, even with no energy resolution, provide some discriminating power.

Further analysis of the prospects for spectral measurements will require more detailed simulations of the radio spectra of detected pulses, and the information they provide on the original particle energy.  It is likely that SKA2, with its capacity for higher-significance detections and thus more precise measurements of the radio spectrum, will be particularly important in this application.

\subsection{Identification of cosmic ray sources}
\label{sec:sources}

The two major UHE cosmic ray detectors, the Pierre Auger Observatory (in the southern hemisphere) and Telescope Array (in the northern hemisphere), have each detected a large-scale localised excess of UHE cosmic rays.  Catalogued events from the Pierre Auger Observatory \citep{abreu2010} show an excess aligned with the nearby active galactic nucleus Centaurus~A, with 13 UHE cosmic rays detected within an 18$^\circ$ radius of this object against an isotropic expectation of 3.2 \citep{abreu2011}, although the significance of the result has been reported to decline in more recent data \citep{revenu2013}.  Similarly, Telescope Array observed 19 UHE cosmic rays against an isotropic expectation of 4.5 originating within 20$^\circ$ of \mbox{R.A.\,$146.7^\circ$} \mbox{Dec.\,$+43.2^\circ$}, although this hotspot does not correspond to any similarly prominent astronomical object \citep{abbasi2014b}.  Increased statistics would allow these two positions to be identified or excluded as sources of UHE cosmic rays.

Taken at face value, and allowing for the relative directional exposure of
each detector, these results imply Centaurus~A to be responsible for 7.4\%
of the all-sky UHE cosmic ray flux, and the hotspot detected by Telescope
Array to be responsible for a further 9.5\%.  Lunar observations to test
these possible sources could be scheduled while the Moon is closest to
them: 10\% of the lunar orbit is within 36$^\circ$ of Centaurus~A, or
within 28$^\circ$ of the hotspot, which would allow 1,000 hours of targeted
observations to be carried out over \mbox{2--3} years.  At these angular
offsets from the Moon, the directional aperture for SKA1-LOW is 5--15 times
the all-sky average aperture, depending on energy, allowing a
correspondingly increased exposure.  Consequently, lunar observations with
SKA1-LOW would expect to detect 5.3 (1.6) cosmic rays originating from
Centaurus~A or 9.4 (3.1) from the hotspot in scenarios~A~(B) considered
above; these numbers are in addition to those listed in
Table~\ref{tab:cr_rates}.  We have neglected here the case of a local void
in cosmic ray sources, because the existence of a single resolvable source
implies that it lies within the local universe.  We have also assumed the
putative source to have the same spectrum as the cosmic ray background,
which is pessimistic, because cosmic rays from a nearby source should be
less affected by GZK attenuation.

From the mean energy and root-mean-square angular deviation of the events detected around Centaurus~A, normal random magnetic deflection of 8.3$^\circ \times (10^{20} {\rm eV} / E)$ would be sufficient to explain the distribution of the excess.  The corresponding figure for the northern-hemisphere hotspot is 9.9$^\circ \times (10^{20} {\rm eV} / E)$.  We use the mean of these two similar values to obtain the empirical representation of the magnetic deflection shown in Fig.~\ref{fig:angres}.  By taking this simple dependence on energy, we have effectively assumed the cosmic ray composition to remain constant in this energy range.

Using the above assumptions, we perform a simple simulation to determine the expected significance of a detection of the hotspot, as the easier to test of the two putative sources.  We apply an increased weight to higher-energy cosmic rays, using as our figure of merit the sum over detected cosmic rays
 \begin{equation}
  M = \sum \theta^{-2} \, \Omega^{-1}
 \end{equation}
where $\theta$ is the angular displacement of a detected cosmic ray from the putative source and
 \begin{equation}
  \Omega = \pi \, (\sigma_{\rm rad}^2 + \sigma_{\rm tan}^2 + \sigma_{\rm mag}^2)
 \end{equation}
is a field of view based on the angular displacements in Fig.~\ref{fig:angres}.  To represent the limited energy resolution, when determining the weighting factor $\Omega$ we randomise the energy with a log-normal distribution with half-width 0.37 in $\ln E$, taking the pessimistic end of the range in Sec.~\ref{sec:energres}.

Performing this simulation for both scenarios considered above, with an isotropic background only, we find that a threshold of \mbox{$M = 5000$~rad$^{-2}$} will be exceeded in \mbox{$<5$\%} of trials in either case, allowing a 95\%-confidence result.  With the hotspot included as described above, this threshold is exceeded in 76\% of trials in scenario~A, but only 16\% of trials with the softer spectrum of scenario~B.  This simplified simulation indicates that the lunar observations considered here have the potential to provide independent confirmation of the hotspot in cosmic rays, but this will depend on the behaviour of their spectrum.  More sophisticated simulations \citep[e.g.\ those of][]{rouille_dorfeuil2014} would find the expected results across a wider range of scenarios.

\subsection{Detection of UHE neutrinos}
\label{sec:neutrinos}

\begin{figure}
 \begin{center}
  \includegraphics[width=0.6\linewidth]{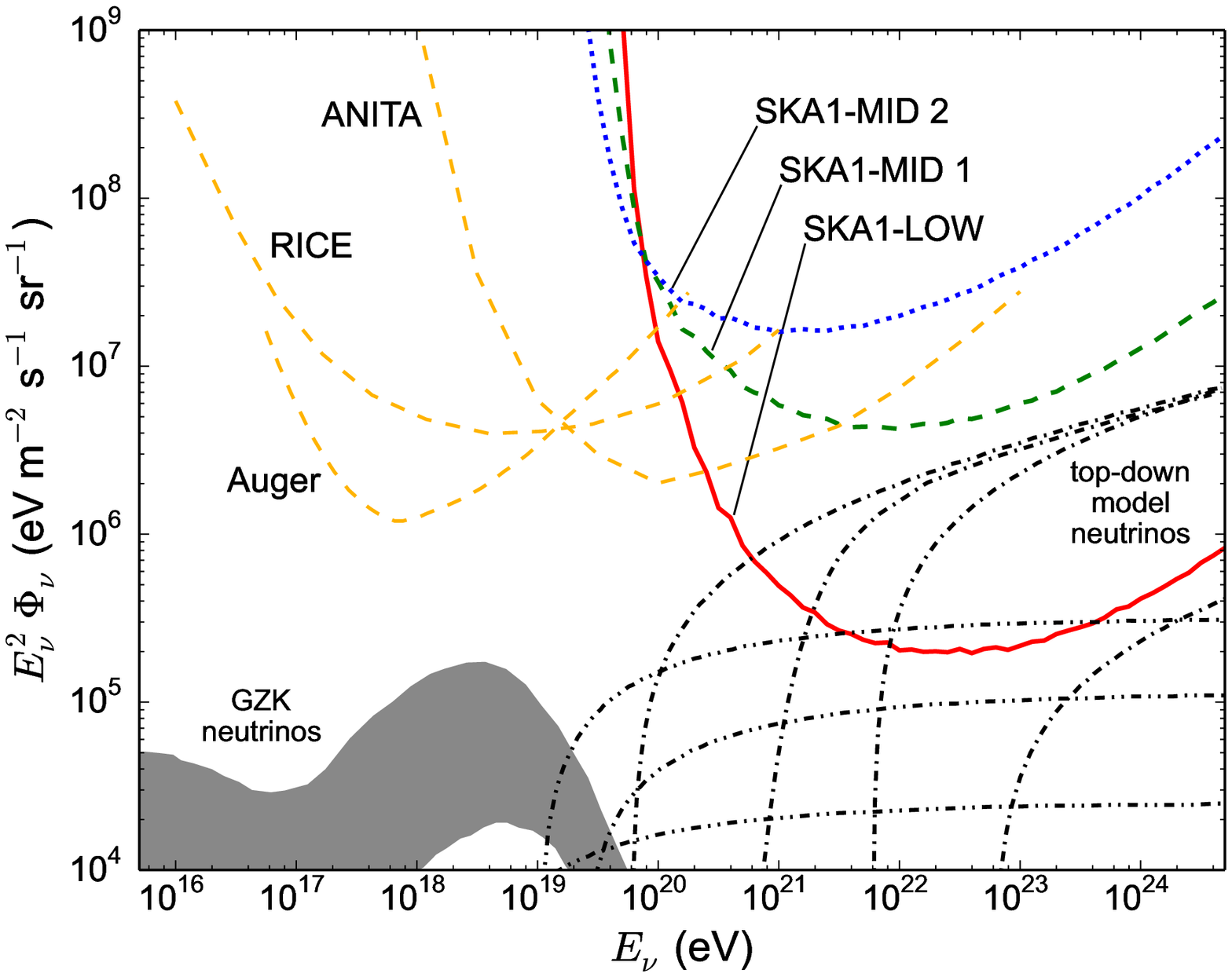}
  \caption{Projected 90\%-confidence limits on the UHE neutrino flux from
1,000 hours of observations with different bands of SKA1.  Predictions are shown for the neutrino flux from GZK interactions \citep[shaded]{Allard06}, and from top-down models involving the production of UHE neutrinos from kinks \citep[dash-dotted]{Lunardini12} and cusps \citep[dot-dash-dotted]{Berezinsky11} in cosmic strings.  Limits set by other experiments --- the Pierre Auger Observatory \citep{AUGERrio}, RICE \citep{RICE} and ANITA \citep{ANITA,ANITAerratum} --- are also shown.}
  \label{fig:nu_lim}
 \end{center}
\end{figure}

UHE neutrinos are expected to be produced by GZK interactions as well as, in exotic top-down models, directly alongside UHE cosmic rays.  Top-down models generally predict much harder neutrino spectra than those from GZK interactions, and are commonly constrained by limits on the fluxes of UHE neutrinos and photons \citep[e.g.][]{ANITA,abraham2009b}.  Families of predicted neutrino spectra for both cases are shown in Fig.~\ref{fig:nu_lim}, alongside existing and projected limits.  No band of SKA1 is sensitive to the GZK neutrino flux, but some top-down models could be detected with SKA1-LOW.  Due to the extreme energies of the neutrinos predicted in this case, they should be clearly distinguishable from the expected lower-energy cosmic rays, so even a single detection could provide strong support for a top-down model.

\subsection{Effects of reduced SKA performance}

The primary characteristic of the SKA which determines its performance in this application is the point-source sensitivity, proportional (for the detection of coherent pulses) to the product of the collecting area and the beamformed bandwidth.  If this sensitivity is decreased by 50\%, the expected detection rates in Table~\ref{tab:cr_rates} are reduced typically by a factor \mbox{$\sim 3$}, requiring a corresponding increase in observing time to compensate.  If the observing time is held at 1,000 hours, the identification of a cosmic ray source considered in Sec.~\ref{sec:sources} is unlikely to be significant, and there is effectively zero capacity to discriminate between the model spectra considered in Sec.~\ref{sec:spectrum}.  Even the probability of achieving a single successful detection of a cosmic ray to verify the technique, as described in Sec.~\ref{sec:verification}, is decreased to \mbox{$\lesssim 80$\%} for a pessimistic cosmic ray spectrum.  The only goal not strongly affected is the testing of exotic top-down neutrino models described in Sec.~\ref{sec:neutrinos}, as the stringency of the limits in this extremely high-energy regime depends more on raw observation time than on radio sensitivity.

\section{Conclusion}

The use of the lunar Askaryan technique offers a new way to study
ultra-high-energy cosmic rays.  Compared to other forms of radio astronomy,
this technique has an unusual set of technical requirements, with
full-time-resolution beamforming, real-time dedispersion and a
sophisticated buffering and triggering algorithm required in order to
detect a nanosecond-scale radio pulse from a particle interaction in the
Moon.  A thorough understanding of these challenges, and the means to
overcome them, has been developed through a series of previous experiments.

SKA1 will be the first radio telescope with sufficient sensitivity to
successfully apply this technique to detect an ultra-high-energy cosmic
ray.  It will also probe a range of astrophysical models: neutrinos from
exotic top-down models, and the flux and anisotropy at the top end of the
cosmic ray spectrum.  Finally, it will develop this technique for use with
SKA2, which will act as an even larger detector for studying
the most extreme particles in the universe.


\setlength{\bibsep}{0.0pt}
\bibliographystyle{apj}

\end{document}